\documentclass[onecolumn,showpacs,nobibnotes,nofootinbib]{revtex4}
\usepackage{amsmath,amssymb}
\usepackage{epsfig}

\def\ba{\begin{eqnarray}}
\def\ea{\end{eqnarray}}
\def\eaa{\end{array}}
\def\beqar{\begin{array}}
\def\bars{\begin{eqnarray*}}
\def\ears{\end{eqnarray*}}

\def\be{\begin{equation}}
\def\ee{\end{equation}}
\def\f{\frac}
\def\D{\delta}
\def\g{\gamma}
\def\ab{\f {\alpha_s N_c}{2\pi^2}}
\def\ac{\f {\alpha_s N_c}{\pi}}

\def\D{\Delta}

\def\G{\Gamma}

\newcommand{\bt}{\begin{tabular}}
\newcommand{\et}{\end{tabular}}
\newcommand{\bd}{\begin{displaymath}}
\newcommand{\ed}{\end{displaymath}\noindent}
\newcommand{\ec}{\end{center}}
\newcommand{\bc}{\begin{center}}

\begin{document}

\title{A scale-invariant ``discrete-time''  Balitsky Kovchegov equation}

\author{A.Bialas%
\footnote{e-mail: {\tt bialas@th.if.uj.edu.pl}}
}
\affiliation{Institute of Physics, Jagellonian University, Reymonta 4,
  30-059 Krakow, Poland.}  
\author{R. Peschanski%
\footnote{e-mail: {\tt pesch@spht.saclay.cea.fr}}
}
\affiliation{Service de physique th{\'e}orique, CEA/Saclay,
  91191 Gif-sur-Yvette cedex, France\footnote{%
URA 2306, unit\'e de recherche associ\'ee au CNRS.}}

\begin{abstract} We consider a version of QCD dipole cascading
corresponding to a finite number  $n$ of discrete $\D Y$ steps of branching
in rapidity. The disctretization scheme preserving the  holomorphic 
factorizability and scale-invariance in position space of the  dipole 
splitting function, we 
derive an exact
recurrence formula from step to step which plays the 
r\^ole of a ``discrete-time'' Balitsky-Kovchegov equation.   The
 BK 
solutions are recovered in the limit $n=\infty$ and $\D Y=0.$ 
\end{abstract}

\maketitle

\section {Introduction} As was first suggested in \cite{Mueller:1993rr}, the
distribution of partons inside a hadron, in the limit of high-energy, and
fixed (large) $Q^2$ and large $N_c$, can be approximated by a cascade of 
colourless
dipoles. Starting from this observation, and using the evolution
equation describing the development of the cascade in rapidity $Y,$ it
was possible to obtain -in the leading logarithmic approximation- an
evolution equation for the scattering amplitude of the dipole cascade on
an ``uncorrelated'' target satisfying the unitarity constraint
\cite{Kovchegov}. The dipole cascading formalism gives a derivation of QCD 
evolution 
equations obtained from the perturbative  expansion in the leading logs 
approximation in energy. Both   
the linear regime
corresponding  to the Balitsky Fadin Kuraev and Lipatov 
(BFKL) 
evolution equation
\cite{Lipatov:1976zz},
and its non-linear extension, the Balitsky-Kovchegov (BK) equation
\cite{Kovchegov,Balitsky:2001gj}, find a convenient description in terms of 
dipoles. 
A general characteristics of the
solutions of the BK equation was recently discussed in
\cite{Munier:2003vc}. It was shown that an approximate geometrical
scaling is the generic asymptotic energy property of the system, related to 
mathematical solutions in terms of traveling waves,  and a
general method of finding these solutions was developped.

The dipole cascading formulation of Refs.\cite{Mueller:1993rr,Kovchegov}
corresponds to a classical branching process where dipoles split with a
probability distribution given by the BFKL 
kernel \cite{Lipatov:1976zz}
expressed in transverse position space taking the role of
(2-dimensional) space, the rapidity variable having the role of {\it
time}. As such it belongs to the large family of random branching
processes which is largely studied in statistical mechanics and in
mathematics (see \cite{FKPP,Derrida:88,bramson,ebert,Munier:2003vc}). While 
the
original QCD dipole formulation of \cite{Mueller:1993rr,Kovchegov} is
based on random branching, we remark that well-known
classes of branching processes are considered with discrete 
steps in
time \cite{Derrida:88}, with interesting physical and mathematical 
properties, and
applications. We want to address here the question of a similar
discretization of dipole cascading.

The attractive feature of the QCD dipole approach is that -being
formulated in the framework of the QCD perturbation theory- it is a
well-defined stochastic fragmentation system. In particular the cascade 
vertices are uniquely
given by the theory. Any departure from this scheme runs into a problem
of serious ambiguities even if one imposes the condition that in the
some limit one should recover the known perturbative results. 

In the present paper we study one class of dicrete-in-rapidity dipole 
cascades  whose
vertices are scale-invariant in transverse space. In particular we retain the 
important 
feature
of {\it holomorphic separability} which is present in the 
leading-logarithmic approximation. We show that such a theory
naturally leads to discretization of the dipole cascade and that in the
 limit of a large number of rapidity steps one 
recovers the results of the leading-logarithmic
approximation. One may hope that this exercise will help to understand
the structure of gluon cascading beyond the lowest order of the
perturbation theory.

The plan of our study is the following. In the next section 2, we derive
a discretized version of the dipole cascading preserving scale-invariance and 
the holomorphic separability of the original BFKL kernel. In section 3 we 
derive the 
corresponding
non-linear master equation, which plays de role of a  ``discrete time'' 
Balitsky-Kovchegov (discrete-time BK) equation. In the next section 4, we use 
the method of  
traveling wave solutions in order to explore the solutions of the 
discrete-time BK 
equation when the
number of steps goes to infinity. In particular we prove 
the convergence to the BK solutions when  the rapidity step  size goes in the 
same proportion to
zero.  In section 5, we draw some possible interesting outcome of our 
discretisation procedure.

\vspace{.5cm}

\section{The master equation} 
\begin{figure} [t]
\epsfig{file=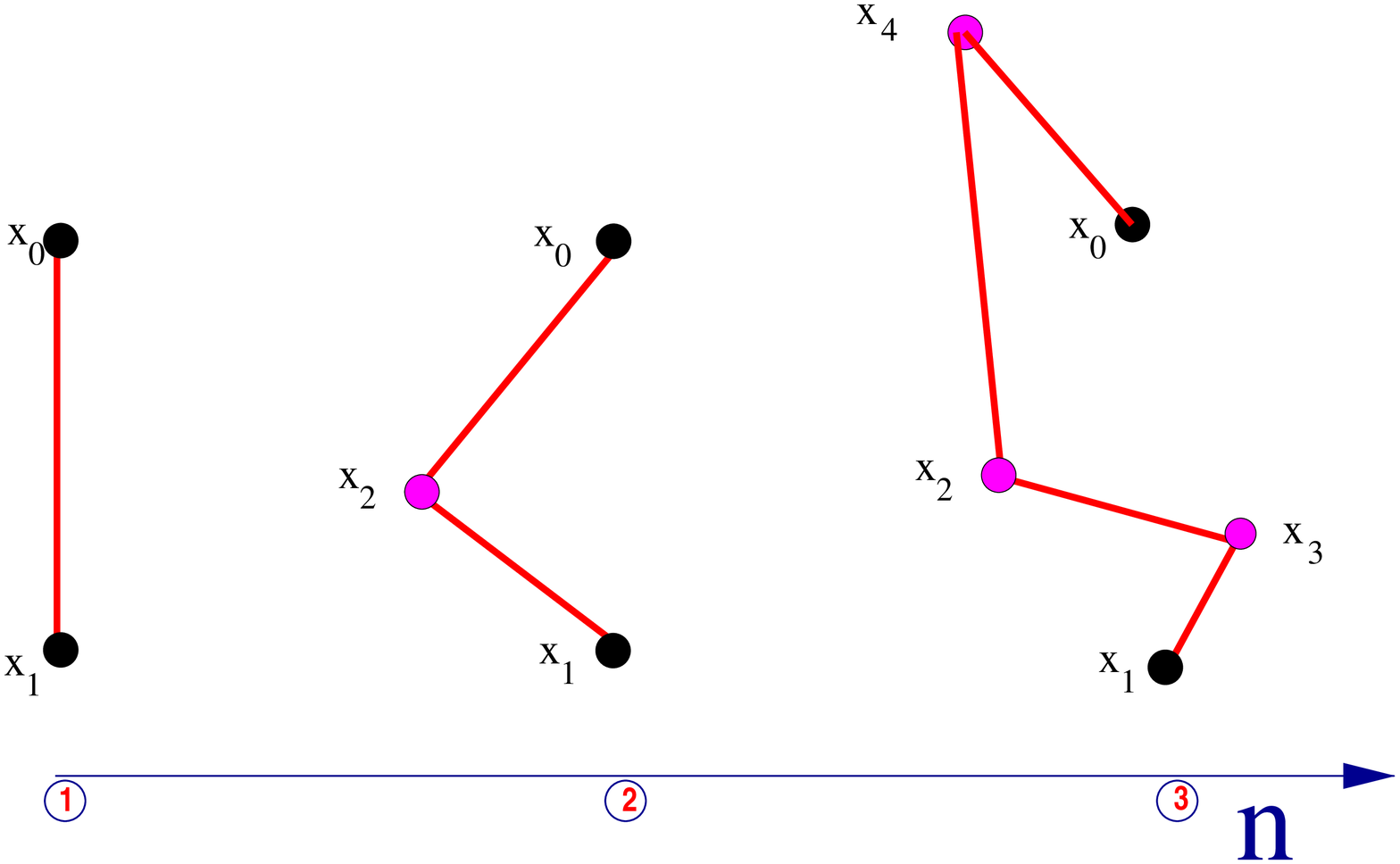,width=12cm}
\caption{\label{f1}{\it Discretized branching process of QCD dipoles.}
The figure shows the rapidity steps for $Y$ evolution of one dipole 
$x_0,x_1$for  
 successive splittings at transverse space points $x_2$ and then $x_3,x_4,$ 
etc...}
\end{figure}
Starting from the the Balitsky-Kovchegov 
equation we  
show 
how the 
modification of the 
vertex leads to the discretization of the dipole cascade.

The BK equation  reads:
\be
\f{d\cal S}{dY}(x_{01},Y) = \int d^2x_2\ {{\cal K}}(x_0,x_1;x_2)\ 
\left\{{\cal 
S}(x_{02},Y) {\cal S}(x_{12},Y)-{\cal S}(x_{01},Y)\right\}\ ,
\label{equation}
\ee
where $S(x_{01},Y)$ is the S-matrix element for the dipole-target amplitude, 
$x_{01}\equiv x_{0}\!-\!x_{1}$ being the dipole transverse size. The BFKL 
kernel 
\be
{{\cal K}}(x_0,x_{1};x_{2})\ d^2x_2 = \ab \ \f {x_{01}^2}{x_{02}^2 x_{12}^2}\ 
d^2x_2\ 
\label{kernel}
\ee
has the classical interpretation \cite{Mueller:1993rr} of  
 a  splitting probability in transverse space by unit of rapidity. 
Note that this kernel verifies simultaneously  scale invariance, the
$x_{02} \leftrightarrow x_{12}$ symmetry and  holomorphic separability. To 
see 
this 
last property, it is worth  using the 2-dimensional components of scaled 
variables  $\xi^{(i)} \equiv  {x_{i}}/{x_{01}}$
and 
transforming 
into complex variables, namely
\ba
\xi^{(1)} + i\xi^{(2)} = z ,\ \xi^{(1)} - i\xi^{(2)} = \bar z\ .
\label{complex}
\ea
 Holomorphic 
separability 
means the factorization property of the kernel (\ref{kernel}) into analytic 
(function 
of $z$) and antianalytic (function of $\bar z$) parts. It plays an important 
role in 
the conformal symetry properties of the BFKL amplitudes \cite{lip}.
 
Let us consider a modified kernel preserving this property:
\be
{\tilde {\cal K}}(x_0,x_{1};x_{2})\ d^2x_2= \ab \ \left(\f 
{x_{01}^4}{x_{02}^2 
x_{12}^2}\right)^{1-a}\ \f {d^2x_2}{x_{01}^2}\equiv \ab \ 
\left(z_{ }\bar z(1\!-\!z)(1\!-\!\bar z)\right)^{a-1}\ \f {dzd\bar z}{2i} ,
\label{kernel-eps}
\ee
and call 
\be
{\cal N} =\int d^2x_2\ {\tilde {\cal K}}(x_0,x_1;x_2)\ ,
\label{N-eps}
\ee
where ${\cal N}$ is finite  for $0<a<1/2$.

We can now safely work separately on the two terms of (\ref{equation}) and 
write
\be
\f 1{{\cal N}}\f{d\cal S}{dY} = \left\{\int d^2x_2\ {{\cal 
K}}_d(x_0,x_1;x_2)\ 
{\cal 
S}(x_{02},Y) {\cal S}(x_{12},Y)\right\}-{\cal S}(x_{01},Y)\ ,
\label{equation1}
\ee
where ${{\cal K}}_d \equiv {\tilde {\cal K}}/{\cal N}$ is now a properly 
normalized 
probability 
distribution.

Using the  known \cite{kawai} mathematical 
identity\footnote{There is one 
condition for (\ref{A3}) to be valid, namely 
$A\!-\!\tilde A,
B\!-\!\tilde B \in {\mathbb Z} ,$ which also ensures that formula
(\ref{A3}) is symetric in the interchange $A\to \tilde A, B\to\tilde B.$}
\be \f 1{2\pi i}\int dzd\bar {z} \ z^{\left(A -1\right)} \bar
{z}^{\left(\tilde A -1\right)} \left(1-z\right)^{\left(B-1\right)}
\left(1-\bar z\right)^{\left(\tilde B -1 \right)} = \frac {\Gamma
\left(A\right)\ \Gamma \left(B\right)\ \Gamma (1- \tilde A- \tilde
B)}{\Gamma (1-\tilde A ) \ \Gamma (1-\tilde B )\ \Gamma
\left(A+B\right)}\ , \label{A3}
 \ee
  we obtain the expression for the normalization
\be
{\cal N} =\frac{\alpha_s N_c}{2\pi}
 \times  \frac {\Gamma ^2 \left(a\right)\ \Gamma  (1- 2a)}
{\Gamma^2  (1-a) \ \Gamma  
\left(2a\right)}\ ,
\label{N-eps=}
\ee
and the following expression for the probability distribution
\be
{{\cal K}}_d (x_0,x_{1};x_{2})\ d^2x_2= \frac1{\pi}\left(\f 
{x_{01}^4}{x_{02}^2 
x_{12}^2}\right)^{1-a}\ \f {d^2x_2}{x_{01}^2}
 \times  \left\{\frac {\Gamma ^2 \left(a\right)\ \Gamma  (1- 2a)}
{\Gamma^2  (1-a) \ \Gamma  
\left(2a\right)}\right\}^{-1}\ .
\label{R}
\ee

Choosing a finite 
interval  
\ba
{\cal N} \Delta Y \equiv \Delta n =1   \label{dy}
\ea
and considering  $n \in 
{\mathbb N}, n \gg 1$ allows one to  transform Eq. (\ref{equation1}) into a 
finite difference   equation
\be
{\cal S}_{n+1}(x_{01})-{\cal S}_{n}(x_{01})= \int d^2x_2\ {{\cal 
K}}_d(x_0,x_1;x_2)
\left\{ {\cal S}_n(x_{02}) {\cal S}_n(x_{12})- {\cal S}_n(x_{01})\right\}\ ,
\label{master}
\ee
leading to (the distribution ${{\cal K}}_d,$ see (\ref{R}), being normalized 
to 1)
\be
{\cal S}_{n+1} (x_{01})=
\int {{\cal K}}_{d}(x_0,x_{1};x_{2})\ d^2x_2
 \times  
{\cal S}_n(x_{02})\ {\cal S}_n(x_{12})\ .
\label{BKcascading}
\ee
As clear enough from its formal structure, Eq.(\ref{BKcascading})  is  a  
``discrete 
time'' version of the  Balitsky-Kovchegov equation (\ref{equation}). It 
remains to 
be proven that it leads back to the BK equation when going to its continuous 
limit, 
which will be defined in the next section.

Indeed, 
Eq.(\ref{BKcascading}) has the typical structure for S-matrix
elements defined for a branching process (or tree structure) with discrete 
steps of 
time evolution. see Fig.\ref{f1}. At each step $n \to n+1$ a dipole of size 
$x_{01}$ splits into
two dipoles of sizes $x_{02},x_{12}$ at the point $x_2$ with a
2-dimensional probability distribution ${{\cal K}}_d(x_0,x_1;x_2).$ In this 
description,  
the formula ({\ref{dy}) determines the length of the ``rapidity veto''
 $\Delta Y = 1/ {\cal N}$. The dependence of $\Delta Y$ on $a$ is
shown in Figure 2.

\section{The recurrence structure in momentum space}

An even  simpler form of equation (\ref{BKcascading}) can be obtained in 
momentum space, when one considers solutions independent of the  
impact-parmeter. It  reveals  even better the 
recurrence structure of the discrete-time BK equation. Using a 2-dimensional 
Fourier transform, one defines
\ba
 {\tilde {\cal S}_n^{(a)}({\bf k})}  \equiv 
 \f
 {\int \frac{d^2x_{01}}{x^2_{01}}\ 
(x^2_{01})^{a}\ e^{i{\bf k\cdot x_{01}}}\  
S_n(x_{01})}
 {\int \frac{d^2x_{01}}{x^2_{01}}\ 
(x^2_{01})^{a}\ e^{i{\bf k\cdot x_{01}}}} = 
\f{\int dx \ J_0(kx)\ S_n(x)\ x^{2a-1}}{\int dx \ J_0(kx)\ x^{2a-1}} ,   
\label{def}
\ea
with 
\be
\int dx \ J_0(kx)\ x^{2a-1}= 2^{2a-1}\ k^{-2a}\f{\G(a)}{\G(1-a)}\ .
\label{1a}
\ee

\noindent working in the context of an impact parameter independent 
framework. Note that 
${\tilde {\cal S}_n^{(a)}({\bf k})},$ can be obtained from  a  
convolution of the S-matrix element in momentum space by the Fourier 
transform of the    the weight 
$(x^2_{01})^{a}$ in the integrand of (\ref{def}).

It is 
straightforward to infer from (\ref{BKcascading}) and (\ref{def}) the 
following 
relation
\ba
 {\tilde {\cal S}_{n+1}^{(2a)}({\bf k})}  =
 \left(\frac {\Gamma(1-a)}{\Gamma \left(a\right)}\right)^2
 \int \frac{d^2x_{02}}{x^2_{12}}\ 
\frac{d^2x_{12}}{x^2_{12}}
(x^2_{02})^{a}\ (x^2_{12})^{a}\ e^{i{\bf k\cdot (x_{02}+x_{21})}}\  {\cal 
S}_n(x_{02})\  {\cal S}_n(x_{12})\ ,   \label{relation}
\ea
where the substitution of the integration variable $x_2\to x_{12},$ with 
Jacobian 
unity, has been performed.

Then, the discrete-time BK equation (\ref{BKcascading}) can be rewritten in a 
particular simple way as 
\ba
  {\tilde {\cal S}_{n+1}^{(2a)}({\bf k})}
= 
\left\{ {\tilde {\cal S}_{n}^{(a)}({\bf k})}\right\}^2\ , 
  \label{final}
\ea
which clearly expresses the nature of the non-linear recurrence relation from 
step to step. this recurrence structure is also clear when expressed  in 
terms 
of transition matrix elements with the same conventional notation  as 
(\ref{def})
\ba
  {\tilde {\cal T}_{n+1}^{(2a)}({\bf k})}
= 
2\ {\tilde {\cal T}_{n}^{(a)}({\bf k})} - 
\left\{{\tilde {\cal T}_{n}^{(a)}({\bf k})}\right\}^2\ .
   \label{finalT}
\ea

\vspace{.5cm} 

\section{High energy limits}
\subsection{Linear regime}
 Let us examine the properties of the master equation   
(\ref{BKcascading}) (also  (\ref{final})).  
Considering first the linearized
form of Eq. (\ref{BKcascading}) near ${\cal S}\sim 1$ we have
\be
{\cal T}_{n+1} (x^2_{01})= \int d^2x_2\ {{\cal K}}_d(x_0,x_1;x_2)\ 
\left\{{\cal 
T}_n(x^2_{02}) + {\cal T}_n(x^2_{12})\right\} \ ,
\label{linearcascading}
\ee
where ${\cal T}\equiv 1-{\cal S}$ is the transition matrix element. 
Using the  complex scaled variables  (\ref{complex}) one 
writes
\be
{\cal T}_{n+1} (t)= 2\!\int dzd\bar {z} \  {{\cal K}}_d(z,\bar z)\ 
{\cal 
T}_n(tz\bar z)  \ ,
\label{linearcascading2}
\ee
where 
\be
{{\cal K}}_d(z,\bar z)=  \left(\frac 
{\Gamma^2  (1-a) \ \Gamma  
\left(2a\right)}{\pi\Gamma ^2 \left(a\right)\ \Gamma  (1- 2a)}\right)\ 
\left\{ z\bar{z}(1\!-\!z)(1\!-\!\bar z) \right\}^{a-1}\ .
\label{linearcascading1}
\ee
and the factor 2 comes from the $z\rightleftarrows 1-z$ symmetry of 
${{\cal K}}_d.$

We now introduce the Mellin transform representation
\be
{\cal T}_{n} (t)= \int_{\cal C} \f {d\g}{2i\pi}\ t^{\g}\ \tilde {\cal T}_{n} 
(\g)\ ,
\label{mellin}
\ee
 and easily derive the recursive relation
\be
\tilde {\cal T}_{n+1} (\g)=\tilde {\cal T}_{n} (\g)\ e^{\ \chi^{(a)}(\g)},
\label{recursive}
\ee
where, using  (\ref{A3}),
\be
{\chi^{(a)}}(\g) \equiv \log \left(2\ \frac {\Gamma  
\left(a+\g\right)\ 
\Gamma  
\left(1-a\right)\ \Gamma  (1- 2a-\g)\ \Gamma  (2a)}
{\Gamma  (a) \ \Gamma  (1-a-\g)\ \Gamma  \left(1-2a\right)\ \Gamma  
\left(2a+\g\right)}\right)\ .
\label{I/N}
\ee
 
 Hence, the solution of the linearized equation is simply 
\be
\tilde {\cal T}_{n} (\g)= \tilde {\cal T}_{0} (\g)\ \exp \left \{n 
\chi^{(a)}(\g)\right\}= \tilde {\cal T}_{0} (\g)\ \exp \left \{{\cal N}  
Y\ \chi^{(a)}(\g)\right\}\ ,
\label{recursive2}
\ee
 where $ Y$ is the total rapidity interval and ${\cal N} $ is given  in 
(\ref{N-eps=}).

Let us now consider the limit $a\to 0.$ A
straightforward algebra gives
\be
{\cal N} \  \chi^{(a)} =  \left\{\f 1a \ \ac  + {\cal O}(a)\right\} 
\times 
 \left\{a \left[ 2\Psi(1)-\Psi(\g)-\Psi(1\!-\!\g)\right] + {\cal 
O}(a^2)\right\}
 = \left[\ac (2\Psi(1)-\Psi(\g)-\Psi(1\!-\!\g))\right] + {\cal O}(a)\ 
,
\label{chi}
\ee 
 where one recognizes in the brackets the BFKL kernel function $\chi(\g)$.
 Hence, from  (\ref{recursive2})
\be
\tilde {\cal T}_{n} (\g)= \tilde {\cal T}_{0} (\g) \exp \left \{n 
a 
 \left[( 2\Psi(1)-\Psi(\g)-\Psi(1-\g)\right]\right\} =  \tilde {\cal T}_{0} 
(\g) 
\ \exp \left \{\ac \chi(\g) \ 
 Y\right\},
\label{recursive1}
\ee
 in which we recover the solution of the  BFKL equation.

\vspace{.5cm}  

\subsection{Non-linear regime} 

Let us now consider, in the spirit of 
\cite{Munier:2003vc} for the  Balitskii-Kovchegov equation,  the
asymptotic solutions of the master equations (\ref{BKcascading}) or  
(\ref{final}) in the continuous limit, i.e. for a   large number $n$ of steps 
of the 
cascade. Asymptotic solutions we are discussing now may be valid
only at really extreme energies. Nevertheless we feel that they may be
of some interest. Indeed, 
as we shall see now by inspecting the general properties of the full 
non-linear 
equations for arbitrary $a$,  the master  equations  (\ref{BKcascading}) (or 
equivalently  (\ref{final})) give the same solutions at large $n$ than the 
Balitsky-Kovchegov equation in the  limit $a\to 0$ or, in more mathematical 
terms, stay in the same universality class. We will also extend  our study to 
another type of continuous limit, namely when $n$ is large but  $a$
is kept 
fixed.

Let us start from the expressions  (\ref{mellin}) and  (\ref{recursive2}) 
giving 
the solution of the linearised problem. One has
\be
{\cal T}_{n} (t)= \int_{\cal C} \f {d\g}{2i\pi}\ t^{\g}\ \tilde {\cal T}_{0} 
(\g)\ \exp \left \{{\cal N}  Y\ \chi^{(a)}(\g)\right\}\ ,
\label{mellin2}
\ee
where $\chi^{(a)}$ has been obtained in (\ref{I/N}). We will now follow  
mathematical arguments used for the solution of discrete 
non-linear equations appearing in a statistical mechanic context \cite 
{Derrida:88}. They are similar  than those used in  \cite{Munier:2003vc} for 
the 
asymptotic solutions of the Balitsky-Kovechegov equations and are part of 
more 
general  mathematical results on non-linear equations, such as for the Fisher 
and 
Kolmogorov, Petrovsky, Piscounov equations \cite {FKPP,bramson}  of ``pulled 
front'' type \cite{ebert}.

Similarly to the arguments developped in  \cite{Munier:2003vc}, the solution  
of 
the non-linear equation equations~(\ref{BKcascading}) or  (\ref{final}) 
are 
traveling waves whose expressions can be obtained starting from the linear 
equation (\ref{mellin2}).

Let us reinterpret  (\ref{mellin2})  as   a linear superposition of waves:
\begin{equation}
{\cal T}_n(t)=\int_{\cal C}
\frac{d\gamma}{2i\pi}\,\tilde {\cal T}_{0} (\g)\ 
\exp \left\{{-\gamma (L_{\text{\tiny WF}}+vn)
+n\ \chi^{(a)}(\gamma)}\right\}\ ,
\label{inicond}
\end{equation}
where $L\equiv \log 1/t$ and   $L_{\text{\tiny WF}}\equiv L-vn$ is  looked 
for as 
the scaling variable determining the    region  moving with  the 
traveling 
 wave front.  $ \chi^{(a)}(\g)$ 
defines the dispersion relation of the linearized equation.
In particular, each partial wave of wave number
$\gamma$ has a {\it phase velocity} 
\begin{equation}
v_{\varphi}(\gamma)=\frac{\chi^{(a)}(\gamma)}{\gamma}
\end{equation} 
whose expression is found by imposing
that the exponential factor in 
Eq.(\ref{inicond}) be  independent of $n$ for $v=v_\varphi(\gamma)$.
By contrast, the {\it group velocity} is defined by the saddle point 
$\g^*$
of the exponential
phase factor
\begin{equation}
v^*=\left.\frac{d\chi^{(a)}}{d\gamma}\right|_{{\g^*}}\equiv v_g\ .
\label{groupvelocity}
\end{equation} 
The key point of the mathematical derivation of the asymptotic solution of 
the 
non-linear equation is that, for appropriate initial conditions 
\cite{Derrida:88,Munier:2003vc}, the critical regime at $\g=\g^*$ is selected 
by 
the non-linear damping.

The velocity $v$ of the front is defined by
\begin{equation}
v^*=\frac{\chi^{(a)}(\g^*)}{\g^*}
=\min_\gamma\frac{\chi^{(a)}(\gamma)}{\gamma}\ ,
\label{frontvelocity}
\end{equation}
where the value of $\g^*$ is determined from the equation
\ba
\left.\frac {d\chi^{(a)}}{d\g} = \frac{\chi^{(a)}(\g)}{\g}\right|_{{\g^*}}\ .  
\label{s1}
\ea   
Indeed, in this case, the group velocity is identical to
the minimum of the phase velocity\footnote{The relation 
$v_g=v_{\varphi},$ in analogy with wave physics, has been written in 
Ref.\cite{GLR}.}.
Now, since 
\ba
n=Y/\Delta Y= Y {\cal N}   \label{s2}
\ea
we can write $v^*n=v_Y  Y$, with
\ba
v_Y = v^*\ {\cal N}.          \label{s3}
\ea 
Reporting these results  in Eq.(\ref{inicond})  gives the dominant term
in the asymptotic expression for the form of the front:
\be
{\cal T}( Y,t)\sim \tilde {\cal T}_{0} (\g^*)\ 
\exp \left\{{-\g^* \log 1/t
+{\cal N} \chi^{(a)}(\g^*) Y}\right\}\ .
\label{inicond2}
\end{equation}
Note that two more ``universal prefactors'' (ie. independent of the initial 
conditions and of the precise form of the non-linear damping terms) can be 
obtained \cite{ebert,Munier:2003vc} in the asymptotic expansion of the 
solution.

From (\ref{inicond2}), it is clear that the same result than for the 
Balitsky-Kovchegov equation will be obtained in the limit $a\to 0,$ due to 
the 
equivalence of the linear kernels and the universality properties due to the 
the non-linearities. This achieves the proof.

Let us consider our high energy limits for  $n \to 
\infty$ with $a \neq 
0, 
 {1/2}$ kept fixed. Some features of the asymptotic solutions are displayed 
in 
Fig.\ref{f2}. As shown in the left part of the figure, the rapidity step  
$\Delta Y$ 
is displayed as a function of $a$ for a given value of $\alpha_s =.1.$ There 
is a 
maximum of the reachable rapidity step for  $a \sim .3$  whose value  depends 
on 
 $\alpha_s.$ At this stage this remains an intringuing feature of our 
scheme of discretization. The right hand part of the figure gives the 
effective 
slope (or ``intercept'') in rapidity, which, starting from the BK value  
stays 
constant till approximately $a = .25$ and then grows. Note that the energy 
momentum 
constraint on the number of steps at given rapidity range  cannot be taken 
into 
account in this continuous limit, as will be discussed in the next section.

\begin{figure} [t]
\epsfig{file=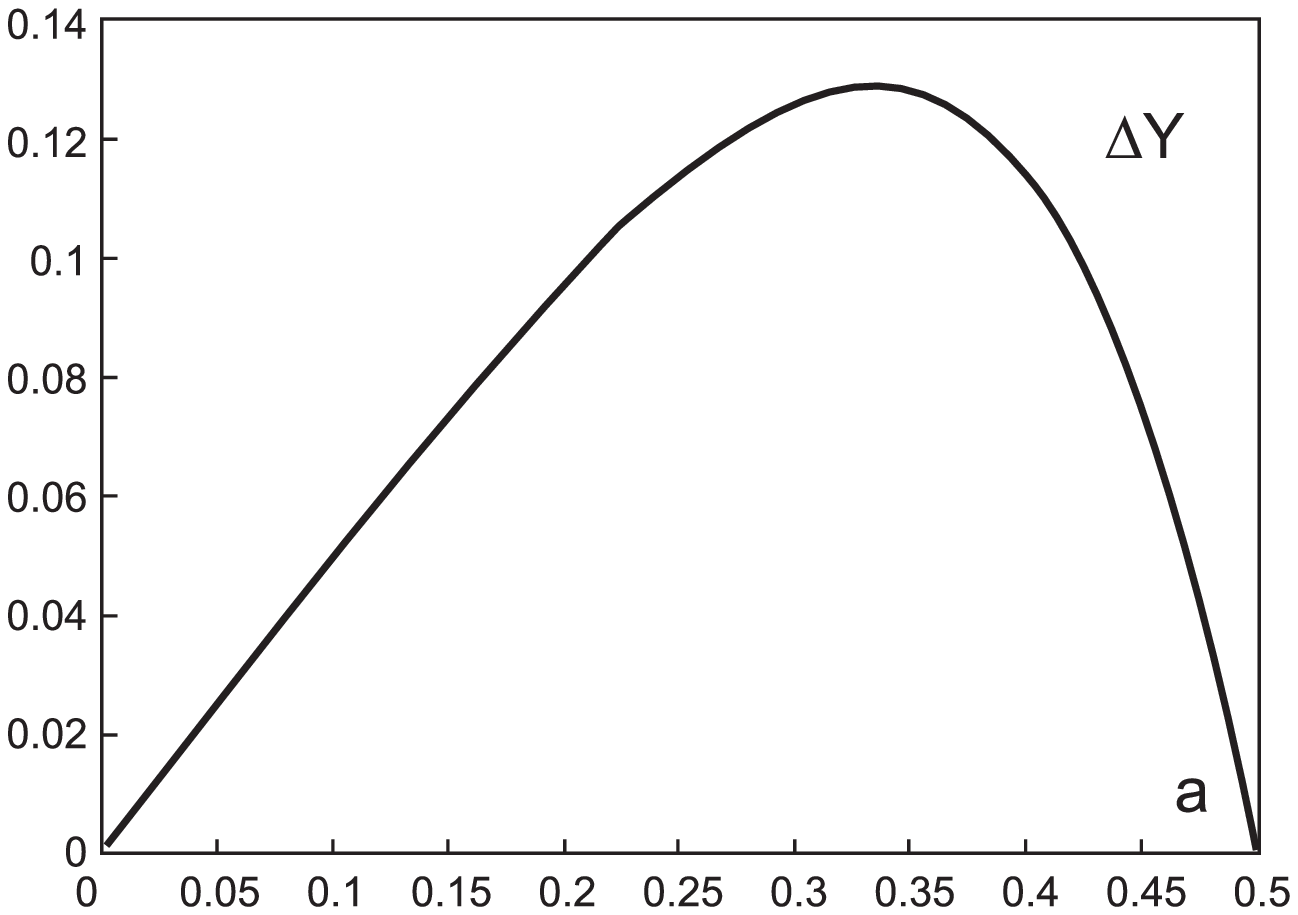,width=8cm}
\hfill \epsfig{file=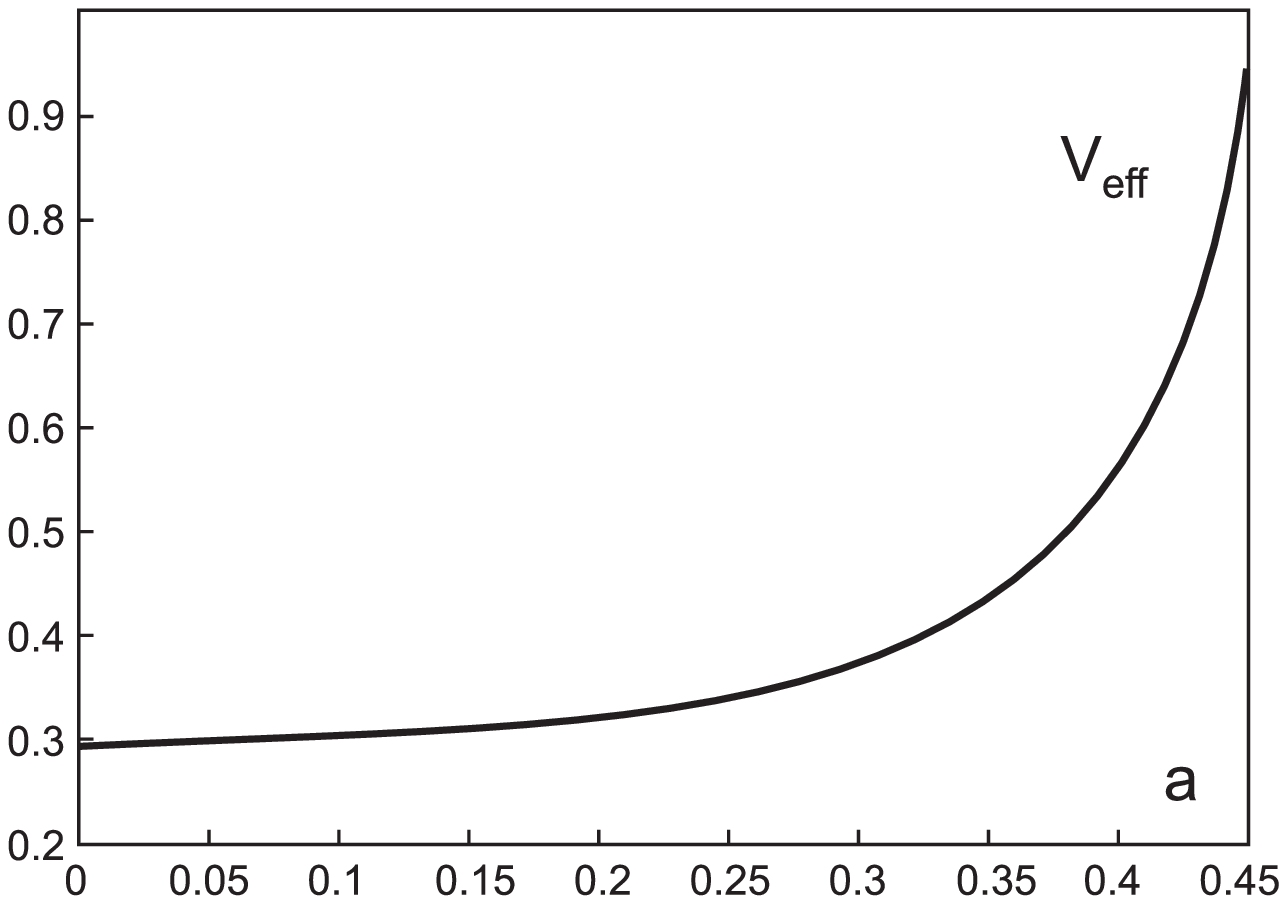,width=8cm}
\caption{{\it Features of the ``discrete time'' solutions.} Left: Effective 
rapidity step $\Delta Y$  in units of $\frac{\alpha_s N_c}{2\pi}$
versus 
$a$.  Right: Effective slope  $v_Y = v^*{\cal N}$ (see formula (\ref{s3})) at  
$\alpha_s=.1$
 versus $a$.}
\label{f2}
\end{figure}
\section{Conclusion and Outlook}
To summarize the main results of our paper, we have considered a ``discrete 
time'' 
version of QCD dipole brancing process characterizing the  approximation of 
the  
perturbative expansion in the leading logs of the energy. This  is obtained 
by 
a 
modification of the splitting probability for finite steps in rapidity 
preserving 
the holomorphic factorizability of the initial BFKL kernel.

Our main result is the derivation of an exact counterpart of the BK 
saturation 
equation  in terms of  recurrence formulae of quite simple form, expressed  
both in 
position space (\ref{BKcascading}) or in a (modified) fourier-transformed 
space 
(\ref{final}).

We checked that we recover the BFKL and BK equations in the appropriate 
$\Delta Y 
\to 0$ continuous limit while the high energy asymptotics for $\Delta Y \ne 
0$ 
have 
been investigated.

The principal outlook of our study concerns the problem of subasymptotics. 
Indeed, 
the master equations    (\ref{BKcascading}) and   
(\ref{final}) 
allow for an iterative solution of the non-linear evolution problem. 
Inserting 
any initial condition ${\cal T}_{0}$ in  (\ref{BKcascading}) or in (modified) 
Fourier 
transform for (\ref{final}), it is possible to generate the full 
solution, at each step of evolution. In particular, as explained in Section 
2, 
for $a \neq 0, 1/2$ the steps of the cascade
are separated by a finite distance $\Delta Y$ in rapidity. Therefore,
for a given total energy, the number of steps in the cascade are limited
by $n_{max} \approx  Y/\Delta Y$, where $ Y$ is the total available
rapidity.  Hence these equations give a convenient way to examine the effect 
of a 
limited number of steps, and thus of dipoles, in the cascading process.
Indeed,  the gluon cascade description of
\cite{Mueller:1993rr,Kovchegov} can only be justified in the leading
logarithmic approximation. The investigations of higher order
corrections shows \cite{Schmidt:1999mz} that they tend to limit the number of 
emitted
gluons\footnote{ This is partly due to the simple
effect of energy conservation which is ignored in the LL
approximation.}. One possibility to take this effect into account is to
forbid the emission of gluons which are too close in rapidity
\cite{Schmidt:1999mz,chacha}. Since the emitted gluons are at the source of 
dipole
splitting \cite{Mueller:1993rr}, it is expected that the formation of
dipoles is similarly limited in rapidity. Such a ``rapidity veto'' can -in
turn- be approximated by a discretized cascade where gluon (or dipole)
emissions is separated by a finite distance in rapidity. We thus feel that 
our 
equations (\ref{BKcascading},\ref{final})) could give an efficient way to 
investigate this problem. It also can be useful to develop the parallel with 
statistical mechanics properties which appeared to be recently fruitful 
\cite{Munier:2003vc}.

When completing the writing of the present work, the paper 
\cite{Kharzeev:2005gn} has appeared, claiming that the solution of an 
evolution with discrete rapidity intervals may lead to a chaotic behaviour. 
It 
stems from a toy model of the BK equation in the form of  a logistic map.
We note that  our equation (\ref{finalT}) has the structure  of a logistic 
map 
if one can neglect the shift $(a)\to (2a)$ in the weight factor. However the 
``Malthusian parameter'' would be   then 2 instead of 3.77 (see the 
discussion 
in Ref.\cite{Kharzeev:2005gn}) and thus is smaller than the value leading to   
chaos. Since our equation  (\ref{finalT}) is a consistent discretization of 
the BK equation, the investigation of its convergence (or not) to the mean 
fied problem  deserves to be studied.

\end{document}